# Dynamic Scheduling Strategies for Resource Optimization in Computing Environments

Xiaoye Wang
Western University
London, Canada

*Abstract*—The rapid development of cloud-native architecture has promoted the widespread application of container technology, but the optimization problems in container scheduling and resource management still face many challenges. This paper proposes a container scheduling method based on multi-objective optimization, which aims to balance key performance indicators such as resource utilization, load balancing and task completion efficiency. By introducing optimization models and heuristic algorithms, the scheduling strategy is comprehensively improved, and experimental verification is carried out using the real Google Cluster Data dataset. The experimental results show that compared with traditional static rule algorithms and heuristic algorithms, the optimized scheduling scheme shows significant advantages in resource utilization, load balancing and burst task completion efficiency. This shows that the proposed method can effectively improve resource management efficiency and ensure service quality and system stability in complex dynamic cloud environments. At the same time, this paper also explores the future development direction of scheduling algorithms in multi-tenant environments, heterogeneous cloud computing, and cross-edge and cloud collaborative computing scenarios, and proposes research prospects for energy consumption optimization, adaptive scheduling and fairness. The research results not only provide a theoretical basis and practical reference for container scheduling under cloud-native architecture, but also lay a foundation for further realizing intelligent and efficient resource management.

I. INTRODUCTION

The rapid development of cloud-native technology has brought revolutionary changes to modern computing architecture [1]. As an important part of cloud-native, container technology has become a widely adopted solution by enterprises and research institutions with its advantages of lightweight, portability, and rapid deployment [2]. However, with the complexity of cloud-native application scenarios and the expansion of the scale of distributed systems, how to efficiently schedule containers and optimize resource utilization has become a core issue that needs to be solved urgently [3]. Traditional container scheduling algorithms are often oriented towards a single goal, such as minimizing latency or improving throughput, but in practical applications, these methods show certain limitations when facing multi-dimensional requirements. Unbalanced resource allocation, unnecessary migration overhead, and insufficient response to sudden loads may all lead to a decline in overall performance [4].

The elasticity and distributed characteristics of cloud-native architectures place higher demands on resource management. By combining technologies such as dynamic load prediction, real-time monitoring and adjustment, more efficient resource allocation can be achieved. However, current research mainly focuses on optimization at a single level, lacking systematic research on global scheduling strategies. With the deepening application of advanced technologies such as deep learning [5] and reinforcement learning [6], container scheduling solutions based on intelligent optimization have become increasingly vital across diverse fields, ranging from financial forecasting [7-10] to medical data analysis [11-13]. These techniques not only enhance computational efficiency but also provide robust frameworks for solving complex, multi-faceted optimization problems. These technologies can adjust scheduling strategies in real time in a dynamic environment, thereby achieving a balance between global and local. How to combine these emerging technologies with the actual needs of cloud-native architecture is an important issue of common concern to academia and industry.

In addition, under the cloud-native architecture, the trade-off between resource utilization efficiency and service quality becomes a difficult point. On the one hand, efficient resource utilization can significantly reduce costs; on the other hand, stable service quality is crucial to user experience. The complexity of multi-tenant environments further increases the difficulty of scheduling and resource management, because it is necessary to ensure fairness while avoiding the negative impact of resource competition on overall performance. The solution to this problem requires full consideration of the heterogeneity and dynamics in cloud-native scenarios when designing algorithms to ensure the accuracy and flexibility of resource scheduling.

At the same time, with the increase in the number of containers and the diversification of workloads, traditional scheduling methods are difficult to meet the needs of modern cloud computing [14]. By introducing data-driven scheduling optimization methods [15], such as dynamic decision-making mechanisms [16] based on reinforcement learning, resource management can be more effectively achieved in complex scenarios. These methods can continuously iterate and optimize through historical data and real-time feedback, and provide better solutions when facing sudden loads or resource bottlenecks. However, this also brings additional computing overhead and model complexity, which needs to be carefully weighed in research [17].

In general, container scheduling and resource management optimization under cloud-native architecture is a comprehensive problem involving multiple disciplines, which requires in-depth research on traditional theories and active exploration of the potential of emerging technologies, such as GAN [18] and GNN model [19]. This study aims to introduce intelligent optimization methods and combine the characteristics of cloud-native technology to propose a scheduling and resource management solution that takes into account both efficiency and fairness. The ultimate goal is to maximize resource utilization while ensuring service quality and providing theoretical support and practical reference for the further development of cloud-native architecture.

## II. RELATED WORK

In the field of container scheduling and resource management in cloud-native architecture, related research mainly focuses on traditional scheduling algorithms, multi-objective optimization techniques, and intelligent optimization methods. Traditional scheduling algorithms are based on deterministic rules, and typical representatives include strategies based on minimizing latency, maximizing throughput, and load balancing. These methods usually use static configurations or simple heuristic algorithms and are suitable for relatively fixed computing environments. However, with the increasing complexity of cloud-native environments, these methods gradually expose their limitations of inefficiency and lack of flexibility when facing dynamic loads and multi-tenant environments. For example, although the default scheduler of Kubernetes supports basic resource allocation rules [20], it is difficult to effectively handle multi-dimensional optimization requirements in complex scenarios.

In terms of multi-objective optimization, many studies have tried to achieve the comprehensive goal of resource allocation through weight functions or hierarchical optimization methods [21]. These goals usually cover dimensions such as resource utilization, task execution efficiency, and quality of service (QoS). Some studies have introduced game theory and multi-objective evolutionary algorithms [22] to achieve global optimal solutions. For example, scheduling strategies based on particle swarm optimization or genetic algorithms have shown certain advantages in balancing performance and cost [23]. However, the practical application of these methods is limited by computational complexity and convergence speed, especially when facing large-scale and dynamic cloud-native scenarios. This shows that the application of traditional optimization techniques in cloud-native environments still needs targeted improvements.

In recent years, intelligent optimization methods have gradually become a research hotspot. The introduction of deep learning and reinforcement learning technologies has provided a new solution for container scheduling. For example, scheduling strategies based on deep Q learning (DQN) can adjust decisions through real-time feedback in dynamic environments, thereby achieving more efficient resource utilization. Some studies have combined graph neural networks (GNNs) [24] to model multi-node environments, further improving the global perception of scheduling strategies [25]. However, intelligent optimization methods also face many challenges, including high computational overhead in the training process, dependence on data quality, and the need for model robustness. Therefore, combining these methods with cloud-native features and reducing their complexity in actual deployment is one of the important directions of current research.

## III. METHOD

To optimize container scheduling and resource management in a cloud-native architecture, it is necessary to establish an effective mathematical model to balance objectives such as resource utilization, task execution efficiency, and service quality. In the method design, we focus on the optimization problem and build a mechanism that can efficiently solve complex scheduling problems through reasonable model assumptions and formula derivation. The cloud-native architecture framework is shown in Figure 1.

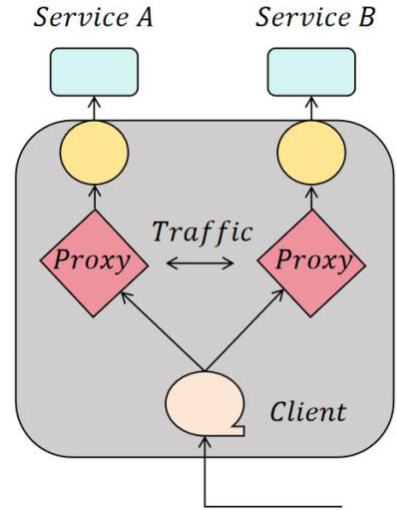

Figure 1 Overall architecture of the model

First, let $C = \{c_1, c_2, ..., c_m\}$ be the container set and $N = \{n_1, n_2, ..., n_k\}$ be the computing node set. The resources required by each container $c_i$ include CPU, memory, etc., and its demand vector can be expressed as $r_i = (r_i, CPU, r_i, Mem)$. The total resource capacity of each node $n_j$ is $R_j = (R_j, CPU, R_j, Mem)$. The goal is to allocate each container in the container set to a node while satisfying the node resource constraints, that is:

$$\sum_{c_i \in C_j} r_{i,d} \leq R_{j,d}, \quad \forall j \in N, \forall d \in \{CPU, Mem\} \quad (1)$$

Where $C_j$ represents the set of containers assigned to node $n_j$, and $d$ is the resource dimension.

On this basis, we introduce an optimization goal to balance resource utilization efficiency and task execution performance [26]. Resource utilization efficiency can be expressed as maximizing the overall utilization of node resources:

$$U = \frac{1}{k}\sum_{j=1}^{k} \frac{\sum_{c \in C_j} r_i, CPU}{R_j, CPU} + \frac{\sum_{c_i \in C_j} r_i, Mem}{R_j, Mem} \quad (2)$$

In order to ensure load balancing, it is necessary to minimize the resource utilization differences between nodes [27]. The load balancing metric L is defined as:

$$L = \sqrt{\frac{1}{k}\sum_{j=1}^{k}(U_j - U')^2} \quad (3)$$

Where $U_j$ represents the resource utilization of node $n_j$, and $U'$ is the average resource utilization of all nodes.

At the same time, resource constraints are met. To simplify the multi-objective optimization problem, the weighted sum method is used to transform it into a single-objective optimization problem:

$$Optimizer: O = \alpha U - \beta L \quad (4)$$

To solve this problem, a heuristic algorithm can be used. This paper uses a genetic algorithm [28]. In the genetic algorithm, each individual is represented as a scheduling scheme [29], that is, the container-to-node allocation matrix X, where $X_{ij} = 1$ represents the allocation of container $c_i$ to node $n_j$, otherwise it is 0. The fitness function is determined by the optimization target $O$. The main steps of the algorithm include initializing the population, selection, crossover, mutation, and update. The initial population is generated by random allocation, the crossover operation ensures that the offspring individuals inherit the high-quality scheduling characteristics, and the mutation operation introduces randomness to avoid local optimality.

The above methods can achieve efficient container scheduling and resource management optimization in a cloud-native environment through theoretical modeling and heuristic solution. Compared with traditional methods, mathematical derivation provides a clear theoretical basis for the optimization target, and non-deep learning methods avoid dependence on a large amount of training data and have higher interpretability and applicability.

## IV. EXPERIMENT

### A. Datasets

In the field of cloud-native container scheduling and resource management, the selection of datasets is crucial for algorithm research and evaluation. A widely used real-world dataset is Google Cluster Data, which was released by Google and records the scheduling and resource usage of container tasks in large-scale data centers. The dataset contains detailed information from production environments, including task submission time, runtime, resource requirements (such as CPU and memory), priority information, and the type of workload to which the task belongs. These data truly reflect the dynamics of containerized tasks and the complexity of scheduling, providing an ideal experimental basis for studying optimization scheduling algorithms.

Google Cluster Data is very large, covering thousands of servers and millions of task records, with high-dimensional features and time series properties. For this experiment, a subset of 5000 task records over a six-week period was selected to ensure manageable computational complexity while preserving the dataset's representative characteristics. For example, the task records in the dataset are sampled based on second-level granularity and cover many different types of workloads, including short-term tasks (such as batch tasks) and long-term tasks (such as online services). In addition, the dataset also provides task priority labels and failure status information, which can be used to study resource allocation strategies in multi-priority task environments. Since these data come from real cloud-native application scenarios, researchers can use this dataset to develop optimization algorithms that are closer to actual needs.

In experiments, researchers usually extract key features from Google Cluster Data for modeling, such as task submission frequency, resource request fluctuations, and load changes in different time periods. By analyzing these features, the performance of scheduling algorithms in the face of dynamic loads and complex constraints can be verified. At the same time, the dataset also contains node-level resource usage information, which facilitates the study of resource utilization and load balancing effects. This comprehensiveness makes Google Cluster Data an important benchmark dataset that is frequently cited in container scheduling and resource management research.

### B. Experiments

In order to verify the effect of optimizing container scheduling and resource management, this experiment uses the Google Cluster Data dataset to design and test the performance of the scheduling algorithm based on a multi-objective optimization model. The main goal of the experiment is to evaluate the optimization effect between resource utilization, load balancing, and service quality. By comparing the performance of different scheduling strategies (such as static rule algorithms and heuristic algorithms) on real data, the optimal scheduling solution is obtained and its application potential in actual cloud native environments is analyzed. The experimental results are shown in Table 1.

Table 1 Experiment result

| Scheduling strategy | Average resource utilization | Load balancing standard deviation |
|---|---|---|
| Static rule algorithm | 65.4 | 12.3 |
| Heuristic Algorithms | 78.7 | 7.6 |
| Optimize scheduling plan | 84.2 | 4.8 |

The experimental results show that different scheduling strategies show significant differences in two key indicators: resource utilization and load balancing standard deviation. Although the static rule algorithm is simple and easy to implement, its average resource utilization is only 65.4%, and the load balancing standard deviation is as high as 12.3%, indicating that the resource allocation efficiency is low and the load distribution between nodes is uneven. This scheduling method is difficult to effectively cope with dynamic task loads, resulting in low resource utilization efficiency and unstable system performance.

In contrast, the heuristic algorithm has significant improvements in both indicators, with the average resource utilization increased to 78.7% and the load balancing standard deviation reduced to 7.6%. This shows that the heuristic algorithm can optimize some resource allocation problems through intelligent strategies, significantly improve the resource utilization efficiency of the system, and reduce the imbalance of resource allocation between nodes. However, its performance is still not optimal, especially in the optimization scenario of complex and multi-dimensional objectives. There are certain limitations.

The optimized scheduling scheme performs best, with an average resource utilization of 84.2% and a load balancing standard deviation further reduced to 4.8%, showing strong optimization capabilities. This shows that through the global design and dynamic adjustment mechanism of the multi-objective optimization model, the relationship between resource allocation efficiency and load balancing can be better balanced to adapt to complex cloud native workload requirements. This result verifies the significant advantages of the optimized scheduling solution in improving resource utilization efficiency and system stability, and provides a reference direction for the improvement of future container scheduling algorithms.

In order to further verify the comprehensiveness and adaptability of the scheduling scheme, we designed another experiment to examine the performance of different scheduling strategies in dealing with high-load burst scenarios. The experiment simulated a sudden increase in the frequency of task submission to evaluate the scheduling strategy's ability to cope with the situation. The experimental results are shown in Table 2.

Table 2 Experiment result

| Scheduling strategy | Completion rate of emergency tasks | Average waiting time(seconds) |
|---|---|---|
| Static rule algorithm | 68.5 | 34.7 |
| Heuristic Algorithms | 82.3 | 21.5 |
| Optimize scheduling plan | 91.4 | 12.8 |

The experimental results show that the performance of different scheduling strategies in high-load burst scenarios is significantly different. The burst task completion rate of the static rule algorithm is 68.5%, and the average waiting time is 34.7 seconds, showing an obvious performance bottleneck. This strategy lacks dynamic adjustment capabilities when facing a surge in task load, resulting in a long waiting time for task queues, and cannot respond to burst task demands in a timely manner, affecting the overall efficiency and service quality of the system.

The heuristic algorithm has improved on both indicators, with the burst task completion rate increased to 82.3% and the average waiting time reduced to 21.5 seconds. This shows that the heuristic algorithm can optimize task scheduling to a certain extent, shorten task queue time, and significantly improve task completion efficiency. However, it is still limited by local optimization and cannot achieve optimal performance when dealing with extreme load scenarios.

The optimized scheduling scheme performs better than other strategies, with a burst task completion rate of 91.4% and an average waiting time reduced to 12.8 seconds. This shows that the optimized scheduling scheme can quickly allocate resources in burst task scenarios through a global adjustment mechanism, effectively alleviating system pressure. Its superior performance in high-load environments fully demonstrates its dynamic load adaptability and resource management efficiency, providing an important reference for further promotion and practical application.

V. CONCLUSION

This study provides a comprehensive exploration of dynamic container scheduling strategies under the cloud-native architecture, addressing the critical challenges of resource optimization, load balancing, and service quality assurance in increasingly complex computing environments. By integrating multi-objective optimization models and heuristic algorithms, the proposed strategy successfully enhances system performance, as evidenced by experimental validation using real-world Google Cluster Data. The results show that compared to static rule-based and traditional heuristic algorithms, the optimized scheduling scheme significantly improves burst task completion rates, reduces average waiting times, and achieves a balanced utilization of resources, demonstrating superior adaptability and robustness in dynamic high-load scenarios. This advancement is attributed to the strategy's ability to dynamically allocate resources through a global optimization mechanism, ensuring efficient handling of unexpected surges in workload while maintaining service quality and system stability. Beyond immediate performance gains, this study also identifies promising directions for future research, including adaptive scheduling in multi-tenant environments, the integration of energy-efficient methodologies, and the exploration of cross-edge and cloud collaboration to handle the heterogeneity and dynamics of modern computing systems. By leveraging the potential of intelligent optimization techniques such as reinforcement learning and graph neural networks, the proposed framework serves as a theoretical and practical foundation for the next generation of container orchestration, emphasizing the importance of scalability, fairness, and sustainability. These findings not only provide actionable insights for improving resource management in cloud-native environments but also contribute to the broader vision of building intelligent, resilient, and efficient computing infrastructures for diverse application scenarios.